\documentclass[pra,amsmath,amssymb,two column]{revtex4}
\usepackage{mathrsfs}
\usepackage{amssymb}

\input epsf.tex
\usepackage{graphicx}
\usepackage{amsthm}

\begin{document}

\title{Quantum key distribution based on quantum dimension and independent devices}
\author{Hong-Wei Li$^{1,2,3}$, Zhen-Qiang Yin$^{1,2}$, Wei Chen$^{1,2a}$, Shuang Wang$^{1,2}$,  Guang-Can Guo$^{1,2}$,  Zheng-Fu Han$^{1,2b}$ }

 \affiliation
 {$^1$ Key Laboratory of Quantum Information,University of Science and Technology of China, Hefei, 230026,
 China\\
 $^2$Synergetic Innovation Center of Quantum Information $\&$ Quantum Physics, University of Science and Technology of China, Hefei, Anhui 230026\\
 $^3$ Zhengzhou Information Science and Technology Institute, Zhengzhou, 450004, China}

 \date{\today}
\begin{abstract}

In this paper, we propose a quantum key distribution (QKD) protocol
based on only a two-dimensional Hilbert space encoding a quantum
system and independent devices between the equipment for state
preparation and measurement. Our protocol is inspired by the fully
device-independent quantum key distribution (FDI-QKD) protocol and
the measurement-device-independent quantum key distribution
(MDI-QKD) protocol. Our protocol only requires the state to be
prepared in the two dimensional Hilbert space, which weakens the
state preparation assumption in the original MDI-QKD protocol. More
interestingly, our protocol can overcome the detection loophole
problem in the FDI-QKD protocol, which greatly limits the
application of FDI-QKD. Hence our protocol can be implemented with
practical optical components.

\end{abstract}
\maketitle

{\it Introduction  - } The unconditional security of a perfect
quantum key distribution (QKD) protocol \cite{bb84} has been proved
by applying entanglement distillation technology and an information
theory approach\cite{sec1,sec2,sec3}. Unfortunately, in practice a
practical QKD system is usually composed of imperfect devices. For
example, a real source emits weak coherent pulses, which contain
multiple photons and will leak the secret key information\cite{pns}.
A wavelength dependent beam splitter may leak the basis information
\cite{wavelength}, since it can be controlled by Eve to apply a
man-in-the-middle attack. More generally, the imperfect device in a
practical QKD system may be controlled by Eve, so that unconditional
security of QKD can not be guaranteed\cite{prsec}.

Since it is difficult to include all possible imperfections in a
security analysis model, fully device-independent quantum key
distribution (FDI-QKD) is a very hot topic since it can defend
against all attacks introduced by imperfect quantum devices
\cite{di1,di2,di3,di4,di5}. The FDI-QKD protocol requires a
violation of the Clauser-Horne-Shimony-Holt (CHSH) \cite{chsh}
inequality between two remote parties Alice and Bob; then
unconditional security of the secret key can be guaranteed by
quantum mechanics and the no-signaling principle. However, in
practical experimental realizations, the quantum channel is lossy
and the detection efficiency is restricted. Thus, the FDI-QKD
protocol is usually vulnerable to a no-fair-sampling attack, which
can introduce the detection loophole problem \cite{loophole} in the
Bell test. To avoid the detection loophole problem caused by the
quantum channel loss, Lim {\it et al. } \cite{lim}
 proposed a FDI-QKD protocol with a local Bell test, which requires Bell tests to be carried out only locally in Alice's
laboratory. But, it can not avoid the detection loophole problem
introduced by the limited detection efficiency.

The most vulnerable device in a practical QKD system is the
single-photon detector, which may be controlled by Eve if she can
apply a light blinding attack \cite{lydersen}. Recently, motivated
by the time-reversed entanglement protocol \cite{mdi3,mdi4}, a
measurement-device-independent quantum key distribution (MDI-QKD)
protocol \cite{mdi1,mdi2} was proposed to avoid the detector side
channel attack. In the MDI-QKD protocol, perfect Bennett-Brasard
1984 (BB84) \cite{bb84} quantum states are prepared on Alice's and
Bob's sides respectively, and then the two photons will be
transmitted to Eve to apply a Bell state projection measurement. By
applying the time-reversed entanglement technique, the perfect
entangled state can be assumed to be prepared on Eve's side, and
then Alice and Bob perform a perfect BB84 state measurement
\cite{bb84}. Correspondingly, the final secret key can be generated
after error correction and privacy amplification. The MDI-QKD
protocol can also be realized in practice, where the actual weak
coherent pulse laser will not weaken the unconditional security of
the key when the decoy state method is applied.

In the spirt of the FDI-QKD protocol, we propose a QKD protocol to
weaken the state preparation assumption in the MDI-QKD protocol.
More precisely, we only require the state to be prepared in the
two-dimensional Hilbert space, and Alice's (Bob's) encoding device
is independent of Eve. In our protocol, the state measurement can be
assumed to be a full black box, while the state preparation can be
assumed to be a black box with a dimension restriction. Similarly to
the security analysis method in the FDI-QKD protocol, we apply the
maximal guessing probability to estimate Eve's information. In the
final security key rate formula, the upper bound of Eve's
information is estimated by the CHSH value violation, while Bob's
information can be calculated by the quantum bit error rate (QBER).
Before proposing our protocol, we will give an introduction to the
FDI-QKD and MDI-QKD protocols in the following.

{\it Fully device-independent quantum key distribution - } The
FDI-QKD protocol considers two remote parties Alice and Bob to share
a secret key, where Alice (Bob) privately chooses a random input
number $x\in\{0,1\}$ ($y\in\{0,1,2\}$) and collects an output
$a\in\{0,1\}$ ( $b\in\{0,1\}$). By considering all of the input and
output random numbers, the data can be divided into two parts. The
first part considers the input data $\{x\in\{0,1\}, y\in\{0,1\}\}$
and the corresponding output data $\{a, b\}$, which can be used for
estimating the CHSH value. While the second part considers the input
data $\{x=0, y=2\}$ and the corresponding output data $\{a, b\}$,
which can be used for estimating the QBER and generating the final
secret key. More precisely, Alice and Bob determine the conditional
probabilities $\{p(a,b|x,y),a,b,x,y\in\{0,1\}\}$ to estimate the
practical CHSH value $g$ as the following equation
\begin{equation}
\begin{array}{lll}
g={\sum}_{a,b,x,y}(-1)^{a+b+xy}p(a,b|x,y),
\end{array}
\end{equation}
where the local hidden variable (LHV) bound of $g$ is $2$, while the
quantum non-local theory guarantee that $g\leq2\sqrt{2}$. From
considerations of quantum mechanics, the conditional probability
value can be given by

\begin{equation}
\begin{array}{lll}
p(a,b|x,y)=Tr(\rho A_{xa}\otimes B_{yb})~~x,y,a,b\in\{0,1\},
\end{array}
\end{equation}
where $\rho $ is the state shared between Alice and Bob,  $A_{xa}$
and $B_{yb}$ are measurement operators with the input parameters
$\{x,y\}$ and the output parameters $\{a,b\}$ respectively. Note
that, projective measurements can be assumed without loss of
generality due to the fact that the quantum dimension has no
restriction. The degree of unpredictability of Alice's measurement
outcome $a$ can be quantified by the maximal guessing probability
(Bob's measurement outcome $b$ can be analyzed similarly)

\begin{equation}
\begin{array}{lll}
p_{guess}(a)= max_a p(a|x)=max_a \sum_bp(a,b|x,y).
\end{array}
\end{equation}
where the second equation use the no-signaling principle. By
applying the FDI-QKD security analysis result given by Masanes {\it
et al.} \cite{di5}, the min-entropy bound of Eve's reduced state
conditioned on Alice's system can be given by the following equation
\begin{equation}
\begin{array}{lll}
 H_{min}(a|x,E)\\
=-log_2 p_{guess}(a)\\
\geq1-log_2(1+\sqrt{2-\frac{g^2}{4}})\equiv f(g).
\end{array}
\end{equation}
From this equation, we can see that Eve cannot get any information
($H_{min}(a|x,E)=1$) when the CHSH value reaches $2\sqrt{2}$. If the
CHSH value can be obtained from the LHV theory, Eve can get all of
the secret information ($H_{min}(a|x,E)=0$). In the previous
analysi, security of the FDI-QKD protocol can be proved without any
other assumptions about the practical devices, thus the quantum
system can be assumed to be prepared in a Hilbert space of arbitrary
dimension. When the quantum state is prepared in two dimensional
Hilbert space \cite{Li3}, the upper bound of the maximal guessing
probability can be given by

\begin{equation}
\begin{array}{lll}
p_{guess}(a) \geq p_{guess}(a)_{2dimensional},
\end{array}
\end{equation}
this inequality can be explained by the fact that the dimension of
Eve's state preparation black box has been restricted to 2; thus she
has restricted information with which to guess Alice's measurement
outcome $a$ compared with the original protocol. Correspondingly,
the lower bound of the min-entropy function can be estimated by
\begin{equation}
\begin{array}{lll}
H_{min}(a|x,E)_{2dimensional} \geq  H_{min}(a|x,E).
\end{array}
\end{equation}

{\it Measurement-device-independent quantum key distribution  - } In
contrast to the previous FDI-QKD protocol, the MDI-QKD protocol can
remove all detector side channel attacks, and has no restrictions of
limited detection efficiency and practical quantum channel losses.
But, the security of the original protocol relies on the assumption
that the legitimate users can perfectly characterize the encoding
systems. The basic idea of the MDI-QKD protocol is to consider that
Alice and Bob to prepare the characterized states, and then the
signals interfere at a 50:50 beam splitter on Eve's side. This is
followed by a polarizing beam splitter, and the signals are
projected into either the horizontal or vertical polarization state.
An appropriate measurement can guarantee the projection into the two
Bell states
$|\psi^-\rangle=\frac{1}{\sqrt{2}}(|HV\rangle-|VH\rangle)$ and
$|\psi^+\rangle=\frac{1}{\sqrt{2}}(|HV\rangle+|VH\rangle)$.
Following the time-reversed entanglement idea, the original MDI-QKD
protocol can be assumed to be the entanglement-based BB84 protocol,
where the prepared state can be guaranteed to be the maximal
entangled state, which will be transmitted to Alice and Bob. Then
Alice and Bob apply a perfectly characterized BB84 state
measurement, and the final secret key can be established after error
correction and privacy amplification. In the entanglement based BB84
protocol, Eve's information about the final secret key can be
estimated from the phase error rate introduced in the quantum. Bob's
uncertainty about Alice's measurement outcome can be directly
calculated from the conditional Shannon entropy function.

Unlike the FDI-QKD protocol, the MDI-QKD protocol has no detection
loophole restriction, which requires Alice and Bob to have almost
perfect state preparation. However, the state preparation may have
some imperfections, which cannot be discovered by the legal parties;
thus the imperfection maybe utilized by Eve to apply an attack. Note
that the security of MDI-QKD has a quantum dimension restriction,
and it can be easily verified that Eve can get all of the secret key
information if a high-dimensional state has been prepared by
Alice(Bob). Thus it is a natural question to ask if MDI-QKD protocol
can generate unconditional security key based only on dimension
restriction\cite{yzq}. Fortunately, the answer is positive if we
consider that Alice (Bob) and Eve share independent devices.

{\it MDI-QKD with independent devices  - } We assume that the state
preparation box on Alice's side has random classical input number
$x$ and hidden variable $\lambda_A$, which can be used to decide
Alice's state preparation $\rho_{x,\lambda_A}$. Similarly, the state
preparation $\sigma_{y,\lambda_B}$ on Bob's side can be controlled
by the hidden variable $\lambda_B$ and input random number $y$.
Unlike in the original MDI-QKD protocol, we do not need to perfectly
characterize the state preparation process, and we also require that
Alice (Bob) and Eve share independent devices, which can be
illustrated by the following equations
\begin{equation}
\begin{array}{lll}
P(\lambda_A|B)=P(\lambda_A|E)=P(\lambda_A),\\
P(\lambda_B|A)=P(\lambda_B|E)=P(\lambda_B),\\
P(\lambda_E|A)=P(\lambda_E|B)=P(\lambda_E),
\end{array}
\end{equation}
where $\lambda_E$ is the hidden variable controlled by Eve's device.
In the independent-devices model \cite{brunner}, we can easily prove
that Alice's (Bob's) state should be prepared using the input number
$x$ ($y$) and hidden variable $\lambda_A$ ($\lambda_B$); neither of
them can be controlled or known by Eve's device. Because of the
independent devices, Eve cannot control or know the state
preparation on Alice and Bob's sides through the hidden variable
$\lambda_E$, and it can easily be verified that the MDI-QKD protocol
has no security if the hidden variable $\lambda_A$ ($\lambda_B$) in
the state preparation black box is controlled by Eve.

We first consider that pure states have been prepared by Alice and
Bob in the two dimensional Hilbert space with given input numbers
$x,y\in\{00,01,10,11\}$,
\begin{equation}
\begin{array}{lll}
Alice:
\{|\phi_{00}\rangle_{\lambda_A},|\phi_{01}\rangle_{\lambda_A},|\phi_{10}\rangle_{\lambda_A},|\phi_{11}\rangle_{\lambda_A}\},\\
Bob:
\{|\phi_{00}\rangle_{\lambda_B},|\phi_{01}\rangle_{\lambda_B},|\phi_{10}\rangle_{\lambda_B},|\phi_{11}\rangle_{\lambda_B}\},\\
\end{array}
\end{equation}
where the state preparation can be assumed to be controlled only by
the input numbers.

More generally, when mixed states have been prepared by Alice and
Bob in the two dimensional Hilbert space with input numbers
$x,y\in\{00,01,10,11\}$, the detailed state preparation sets on
Alice's and Bob's sides can be given by

\begin{equation}
\begin{array}{lll}
Alice:
\{\sum_{\lambda_A}\sqrt{p_{\lambda_A}}|\phi_{00}\rangle_{\lambda_A}|\lambda_A\rangle,\\
~~~~~~~~~~\sum_{\lambda_A}\sqrt{p_{\lambda_A}}|\phi_{01}\rangle_{\lambda_A}|\lambda_A\rangle,\\
~~~~~~~~~~\sum_{\lambda_A}\sqrt{p_{\lambda_A}}|\phi_{10}\rangle_{\lambda_A}|\lambda_A\rangle,\\
~~~~~~~~~~\sum_{\lambda_A}\sqrt{p_{\lambda_A}}|\phi_{11}\rangle_{\lambda_A}|\lambda_A\rangle\},\\
Bob:
\{\sum_{\lambda_B}\sqrt{p_{\lambda_B}}|\phi_{00}\rangle_{\lambda_B}|\lambda_B\rangle,\\
~~~~~~~~~\sum_{\lambda_B}\sqrt{p_{\lambda_B}}|\phi_{01}\rangle_{\lambda_B}|\lambda_B\rangle,\\
~~~~~~~~~\sum_{\lambda_B}\sqrt{p_{\lambda_B}}|\phi_{10}\rangle_{\lambda_B}|\lambda_B\rangle,\\
~~~~~~~~~\sum_{\lambda_B}\sqrt{p_{\lambda_B}}|\phi_{11}\rangle_{\lambda_B}|\lambda_B\rangle\},\\
\end{array}
\end{equation}
where $\sum_{\lambda_A} p_{\lambda_A}=\sum_{\lambda_B}
p_{\lambda_B}=1$. The mixed state is prepared by considering
different hidden variables $\lambda_A$ and $\lambda_B$. For example,
if Alice receives the input number 00, the mixed state
$\sum_{\lambda_A}
p_{\lambda_A}|\phi_{00}\rangle\langle\phi_{00}|_{\lambda_A}$ will be
transmitted to Eve correspondingly.

To relax the state preparation limitation in the MDI-QKD protocol,
we propose our protocol as shown in Fig. 1. Note that we assume the
state preparation and measurement boxes are independent, which is
reasonable in practical experimental realization. We must take this
 assumption since Eve cannot distinguish any
pure state $|\phi\rangle_{\lambda_A}$($|\phi\rangle_{\lambda_B}$)
from Alice's (Bob's) encoding ensembles via the LHV $\lambda_A$
($\lambda_B$); thus any pure state $|\phi\rangle_{\lambda_A}$
($|\phi\rangle_{\lambda_B}$) will be treated equally by Eve.

\begin{figure}[!h]\center
\resizebox{9cm}{!}{
\includegraphics{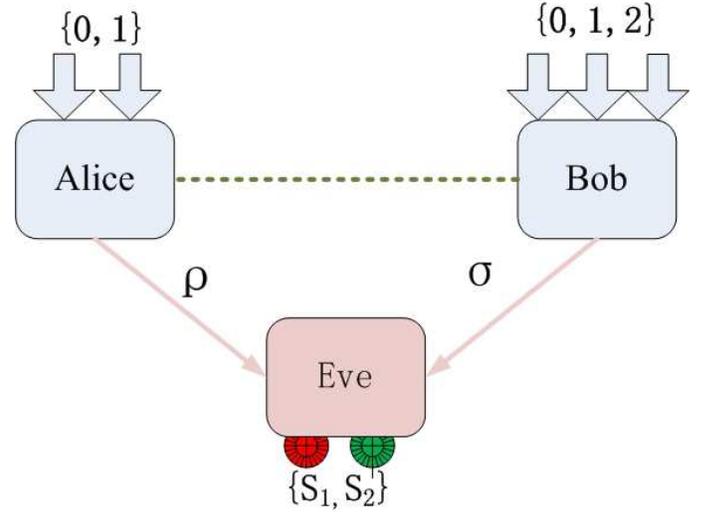}}
\caption{Quantum key distribution based on the quantum dimension and
independent devices}
\end{figure}

Our protocol is illustrated in Fig.1. The detailed steps are
described as follows:

{\it Step 1. State preparation:} Alice prepares two sets of quantum
states $\{\rho_{00},\rho_{01}\}$  and $\{\rho_{10},\rho_{11}\}$ in
the two dimensional Hilbert space, then she randomly chooses one of
the quantum states, which will be sent to Eve in the middle of the
quantum channel. Similarly, Bob prepares three sets of quantum
states $\{\sigma_{00},\sigma_{01}\}$, $\{\sigma_{10},\sigma_{11}\}$
and $\{\sigma_{20},\sigma_{21}\}$, then he randomly chooses one of
the quantum states, which will be sent to Eve in the middle of the
quantum channel.

Without loss of generality, in practical experimental realization,
we can assume that $\{\rho_{00},\rho_{01}\}$ are eigenstates of the
 Pauli operator matrix $Z$, and $\{\rho_{00},\rho_{01}\}$ are
eigenstates of the Pauli operator matrix $X$.
$\{\sigma_{00},\sigma_{01}\}$ are eigenstates of the operator matrix
$\frac{-Z-X}{\sqrt{2}}$, $\{\sigma_{10},\sigma_{11}\}$ are
eigenstates of the operator matrix $\frac{Z-X}{\sqrt{2}}$, and
$\{\sigma_{20},\sigma_{21}\}$ are eigenstates of the pauli operator
matrix $Z$.

 {\it
Step 2. State measurement:} By considering all of Alice's state
preparation sets and Bob's first two sets of state preparation,
Alice and Bob save the classical data when Eve gets the measurement
corresponding to the projection into the Bell state
$|\psi^+\rangle=\frac{1}{\sqrt{2}}(|HV\rangle-|VH\rangle)$. The
measurement results will be noted as the set $S_1$.

Similarly, by considering Alice's first
 state preparation set and Bob's third state preparation set, Alice
and Bob save the classical data when Eve gets the measurement
corresponding to the projection into the Bell states
$|\psi^+\rangle=\frac{1}{\sqrt{2}}(|HV\rangle+|VH\rangle)$ and
$|\psi^-\rangle=\frac{1}{\sqrt{2}}(|HV\rangle-|VH\rangle)$. The
measurement result will be noted as set $S_2$. Note that Bob should
flip his bit value, so that the classical bit 0 (1) will be changed
to bit 1 (0).

 {\it  Step 3. CHSH and QBER value estimation:} Alice and Bob apply
the set $S_1$ to estimate the CHSH value $g$, and the QBER value $e$
can also be estimated by applying the set $S_2$.

{\it Step 4. Error correction and privacy amplification:} By
applying an error correction code, Alice and Bob can establish an
identical classical binary number to eliminate the bit error. Since
Eve can get secret key information from the error correction step
and the non-maximally violated CHSH value, Alice and Bob construct
the final secret key by applying a privacy amplification protocol.

{\it Security analysis model and final secret key rate  - } To
analyze our protocol, we can assume it to be realized in the
following way (Bob's state preparation can be analyzed similarly),
Alice first prepares a pair of systems in the singlet state. If she
wishes to prepare state $|\psi\rangle$, she will measure one
particle in the basis $\{|\psi\rangle,|\psi^\perp\rangle\}$, and the
other particle will also collapse to one of these states. Based on
the measurement outcomes, the second quantum state will be sent to
Eve if Alice gets the measurement outcome $|\psi\rangle$, while the
singlet state will be discarded if Alice gets the measurement
outcome $|\psi^\perp\rangle$. These two cases are demarcated by
fragments (1) and (2) in Fig. 2. The quantum state prepared with the
two different protocols cannot be distinguished, thus security of
the two protocols is equivalent,

\begin{figure}[!h]\center
\resizebox{7cm}{!}{
\includegraphics{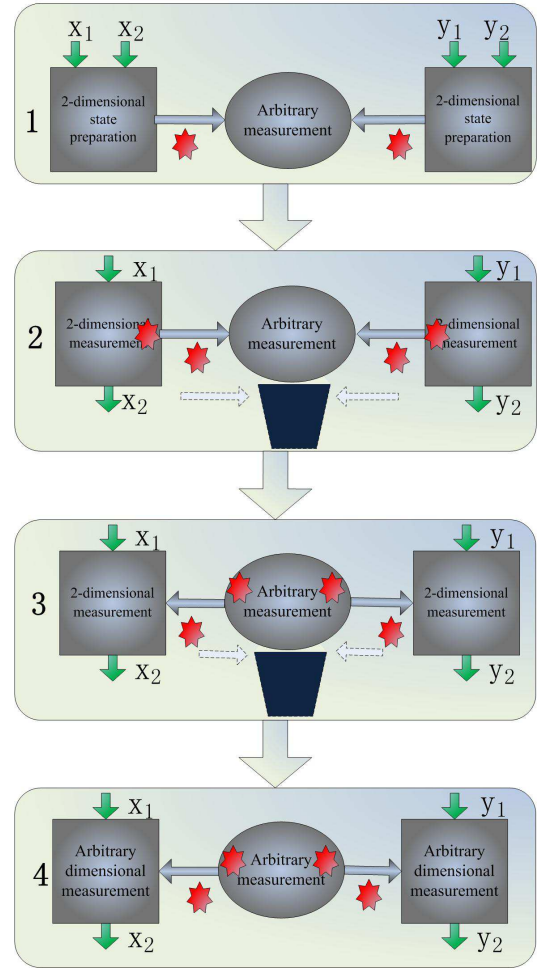}}
\caption{The relationship between the MDI-QKD protocol and the
DI-QKD protocol. In fragments (2) and (3), the singlet state will be
discarded to garbage when Alice or Bob gets the unmatched
measurement outcome. In the MDI-QKD protocol, Alice(Bob) has the
input number $x=(x_1,x_2)$($y=(y_1,y_2)$), which can be transformed
into input number $x_1$($y_1$) and measurement outcome $x_2$($y_2$)
in the DI-QKD protocol. }
\end{figure}

Without loss of generality, we can assume that the singlet states
are prepared by Eve in fragment (3), then she will send one of these
states to Alice. By applying the corresponding two-dimensional state
measurement, Alice informs Eve that she should save the measurement
outcomes if she gets the quantum state $|\psi\rangle$. Note that Eve
is more powerful than in the previous protocol, thus security of the
original protocol is not weakened in the present protocol. Next we
transform the protocol in fragment (3) to the one in fragment (4).
By considering the state measurement equipment as a black box, we
have the protocol shown in fragment (4). Note that Eve's ability
will be enhanced in fragment (4). Obviously, this protocol in
fragment (4) can be assumed to be a DI-QKD protocol, in which we can
apply the CHSH value to estimate Eve's information.

The main difficulty in this work is to obtain the final secret key
rate. We first calculate the final key rate with pure state
preparation on Alice's and Bob's sides respectively. Combining the
CHSH value $g$ with the QBER value $e$, we can get the final secret
key rate $R$ from the following formula

\begin{equation}
\begin{array}{lll}
R\geq H_{min}(a|x,\lambda_A,\lambda_B,E)_{2dimension}-H(a|\bar{b},\lambda_A,\lambda_B)_{S_2}\\
~~\geq H_{min}(a|x,\lambda_A,\lambda_B,E)-h(e)\\
~~=H_{min}(a|x)-h(e)\\
~~\geq1-log_2(1+\sqrt{2-\frac{g^2}{4}})-h(e),
\end{array}
\end{equation}
where $h(x)=-xlog_2(x)-(1-x)log_2(1-x)$ is the binary entropy
function, $\bar{b}$ is the bit flip value of $b$, the second
inequality can be obtained by utilizing the formula (6). Since the
state preparation can be controlled only by the input random number,
the third equation can be proved simply by considering the formula
(8).

As in the pure state preparation case, the final secret key rate
with mixed state prepation can be given by

\begin{equation}
\begin{array}{lll}
R\geq \sum p_{\lambda_A} p_{\lambda_B}
\{H_{min}(a|x,\lambda_A,\lambda_B,E)_{2dimension}\\
~~~~~~~~~~~~~~~~~~~~~~-H(a|\bar{b},\lambda_A,\lambda_B)_{S_2}\}\\
~~\geq \sum p_{\lambda_A} p_{\lambda_B}\{H_{min}(a|x,\lambda_A,\lambda_B,E)-H(a|\bar{b},\lambda_A,\lambda_B)_{S_2}\}\\
~~\geq\sum p_{\lambda_A} p_{\lambda_B}\{-log_2 p_{guess}(a_{\lambda_A,\lambda_B})-h(e_{\lambda_A,\lambda_B})\}\\
~~\geq \sum p_{\lambda_A}
p_{\lambda_B}f(g_{\lambda_A,\lambda_B})-h(\sum p_{\lambda_A} p_{\lambda_B}e_{\lambda_A,\lambda_B})\\
~~\geq f(\sum p_{\lambda_A}
p_{\lambda_B}g_{\lambda_A,\lambda_B})-h(\sum p_{\lambda_A} p_{\lambda_B}e_{\lambda_A,\lambda_B})\\
~~=f(g)-h(e)\\
 ~~=1-log_2(1+\sqrt{2-\frac{g^2}{4}})-h(e),
\end{array}
\end{equation}
where $g_{\lambda_A,\lambda_B}$ and $e_{\lambda_A,\lambda_B}$ are
the CHSH and QBER values with given hidden variables $\lambda_A$ and
$\lambda_B$ respectively. In practical experimental realizations,
the observed CHSH value is $g=\sum p_{\lambda_A}
p_{\lambda_B}g_{\lambda_A,\lambda_B}$, and the observed QBER value
is $e=\sum p_{\lambda_A} p_{\lambda_B}e_{\lambda_A,\lambda_B}$. The
fourth inequality is based on the concave function $h(e)$, and the
fifth inequality is based on the convex function $f(g)$.

We calculate the final secret key rate with different QBER and CHSH
inequality values in Fig. 3.

\begin{figure}[!h]\center
\resizebox{9cm}{!}{
\includegraphics{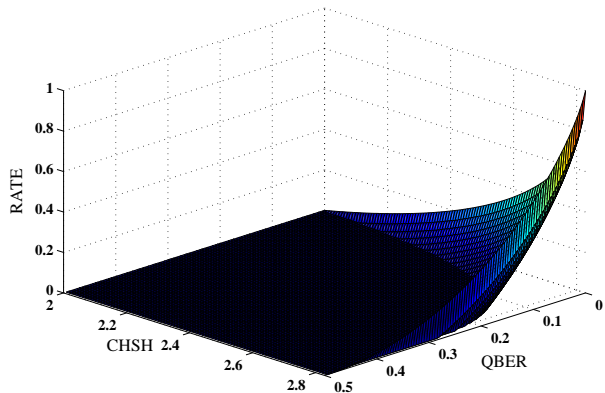}}
\caption{Final secret key rate with different
 QBER and CHSH value }
\end{figure}

From the calculation result, we can see that the maximal tolerable
QBER can reach nearly $0.5$ if the CHSH value reaches $2\sqrt{2}$,
while the maximal tolerable QBER is $0$ if the CHSH value can be
explained by the LHV theory. The calculation result shows that this
protocol is much more robust than the previous MDI-QKD protocol when
$g$ approaches a high value.

In our security analysis, we assume that the state prepared on
Alice's (Bob's) side has no correlation with the previous or the
following state. Where the prepared quantum states can be assumed to
be uncorrelated with each other, the eavesdropper and hidden
variables have no memory in Alice and Bob's devices, and thus the
quantum de Finetti theorem \cite{def} can be directly applied to
make our protocol secure against the most general attack. Other
applications of this protocol and a more general security analysis
are interesting open questions for future research.

{\it Conclusion  - }
 We propose a QKD protocol, the security of
which is based only on quantum states prepared in the two
dimensional Hilbert space and the independent of devices between
Alice (Bob) and Eve. Our protocol can also be practically realized
with current experimental technology.

{\it Acknowledgements  - }
 We thank Chunmei Zhang and Professor Qingyu Cai for
their helpful discussions. H.-W.L. and Z.-Q.Y. controlled equally to
this work, the authors are supported by the the National Natural
Science Foundation of China (Grant Nos. 61101137, 61201239, 61205118
and 11304397), National Basic Research Program of China(Grant No.
2013CB338002) and China Postdoctoral Science Foundation (Grant No.
2013M540514). $^a$weich@ustc.edu.cn, $^b$zfhan@ustc.edu.cn.


\begin{thebibliography}{10}


\bibitem{bb84}  C. H. Bennett and G. Brassard, Proceedings of IEEE International
Conference on Computers, Systems and Signal Processing, Bangalore,
India. New York: IEEE, 1984. 175¨C179

\bibitem{sec1} H.-K. Lo and H. F. Chau, Science 283, 2050 (1999).
\bibitem{sec2}P.W. Shor and J.
Preskill, Phys. Rev. Lett. 85, 441 (2000).
\bibitem{sec3}R. Renner, PhD thesis, Diss. ETH No 16242,
quant-ph/0512258 (2005).


\bibitem{pns}G. Brassard, N. Lutkenhaus, T. Mor, and B. C. Sanders, Physical Review Letters, 85(6):1330
(2000).

\bibitem{wavelength}H. W. Li, S. Wang, J. Z. Huang, et al., Phys. Rev. A 84(6): 062308 (2011)

\bibitem{prsec}V. Scarani, et al.,  Reviews of modern physics. 81(3): p. 1301 (2009).


\bibitem{di1} S. Pironio, A. Acý´n, N. Brunner, N. Gisin, S. Massar, and V.
Scarani, New J. Phys. 11, 045021 (2009).
\bibitem{di2} M. McKague, New J. Phys. 11, 103037
(2009).
\bibitem{di3} E. Haggi, Ph.D. thesis, ETH Zurich, 2010,
arXiv:1012.3878.
\bibitem{di4} E. Haggi and R. Renner,
arXiv:1009.1833.
\bibitem{di5} L. Masanes, S. Pironio, and A. Acin, Nat. Commun. 2, 238 (2011).

\bibitem{chsh} J. F. Clauser, M.A. Horne, A. Shimony, and R. A. Holt, Phys. Rev. Lett.
23, 880 (1969).

\bibitem{loophole} Anupam Garg, N.D. Mermin, Phys. Rev. D 25 (12): 3831¨C5
(1987).

\bibitem{lim}Charles Ci Wen Lim, Christopher Portmann, Marco Tomamiche, Renato
Renner, and Nicolas Gisin, Phys. Rev. X 3, 031006 (2013)

\bibitem{lydersen}Lydersen L, Wiechers C, Wittmann C, et al. Hacking commercial
quantum cryptography systems by tailored bright illumination. Nat
Photonics, 2010, 4: 686¨C689

\bibitem{mdi1} S. L. Braunstein and S. Pirandola, Phys. Rev. Lett. 108, 130502 (2012).
\bibitem{mdi2} H.-K.
Lo, M. Curty, and B. Qi, Phys. Rev. Lett. 108, 130503 (2012).



\bibitem{mdi3}E. Biham, B. Huttner, and T. Mor, Phys. Rev. A 54, 2651 (1996).
\bibitem{mdi4} H.
Inamori, Algorithmica 34, 340 (2002).

\bibitem{Li3} H-W Li, P. Mironowicz, M. Paw{\l}owski, Z-Q Yin, Y-C Wu, S. Wang, W. Chen,
H-G Hu, G-C Guo, and Z-F Han, Phys. Rev. A {\bf 87}, 020302(R),
(2013)

\bibitem{yzq}Z-Q. Yin, C-H. F. Fung, X-F. Ma, C-M. Zhang, H-W. Li, W. Chen, S. Wang, G-C. Guo, Z-F. Han, Phys. Rev. A 88, 062322 (2013).

\bibitem{brunner} J. Bowles, M. T. Quintino, N. Brunner, e-print arXiv:1311.1525
(2013).

\bibitem{def}M. Christandl, R. Konig, and R. Renner, Phys. Rev. Lett. 102, 020504
(2009).

\end{thebibliography}
\end{document}